\documentclass{aa}     
\usepackage{epsfig}
\def\kms{km~s$^{-1}$ }
\def\a85{ABCG~85{}}
\def\bj{b$_{\rm J}${}}
\thesaurus{11.03.4 ABCG~85; 11.03.1 }

\title{A catalogue of velocities in the cluster of galaxies Abell 85
\thanks{Based on observations collected at the European Southern Observatory,
La Silla, Chile}}

 \author {
  F.~Durret \inst{1,2}
\and
  P.~Felenbok \inst{2}
\and
  C.~Lobo \inst{1,3}
\and
  E.~Slezak \inst{4}
}
\offprints{F.~Durret, durret@iap.fr }
\institute{
  Institut d'Astrophysique de Paris, CNRS, Universit\'e Pierre et Marie Curie, 
  98bis Bd Arago, F-75014 Paris, France 
\and 
    DAEC, Observatoire de Paris, Universit\'e Paris VII, CNRS (UA 173),
    F-92195 Meudon Cedex, France 
\and
    Centro de Astrof\'\i sica da Universidade do Porto, Rua do Campo Alegre 
    823, 4150 Porto, Portugal 
\and
    Observatoire de la C\^ote d'Azur, B.P. 229, F-06304 Nice Cedex 4, France 
}
\date{Received, 1997; accepted,}
\begin{document}

\maketitle

\begin{abstract}
We present a catalogue of velocities for 551 galaxies (and give the
coordinates of 39 stars misclassified as galaxies in our photometric
plate catalogue) in a region covering about 100'$\times$100' 
(0.94$\times$0.94~Mpc for an average redshift of 0.0555, assuming 
H$_\circ$=50~km~s$^{-1}$~Mpc$^{-1}$) in the direction of the rich cluster 
\a85. This catalogue includes previously published redshifts by Beers et al.  
(1991) and Malumuth et al. (1992), together with our 367 new measurements. 
A total of 305 galaxies have
velocities in the interval 13350-20000~km~s$^{-1}$, and will be considered as
members of the cluster.  \a85 therefore becomes one of the clusters
with the highest number of measured redshifts; its optical properties
are being investigated in a companion paper.\\ 
\end{abstract}

\section{Observations and data reduction} 

\subsection{Description of the observations}

The observations were performed with the ESO 3.6m telescope equipped with MEFOS
(see description below) during 6 nights on November 5-11, 1994 and 2 nights 
on November 24-26, 1995. The grating used with the Boller \& Chivens 
spectrograph had 300 grooves/mm, giving a dispersion of 224~\AA /mm in the 
wavelength region 3820-6100~\AA . The detector was CCD \#32, with 512$^2$ 
pixels of $27 \times 27\ \mu$m. 

The catalogue of galaxy positions used in this survey was obtained with the 
MAMA measuring machine and is presented in a companion paper (Slezak et al., 
1997). This catalogue gives approximate magnitudes in the \bj \ band,
which were used to select the galaxies to be observed spectroscopically.
CCD photometry of the central regions of the cluster in the V and R bands was 
later performed to recalibrate \bj \ magnitudes and obtain V and R magnitudes
for the entire photometric sample. We observed spectroscopically a total 
number of 21 fields, with exposure times of 2$\times$20 minutes for the two 
fields with galaxies all brighter than b$_{\rm J}$=18, and 2$\times$30 minutes 
for the other ones.  
We obtained 519 spectra in total (plus the same number of sky spectra).

\begin{figure}
\centerline{\psfig{figure=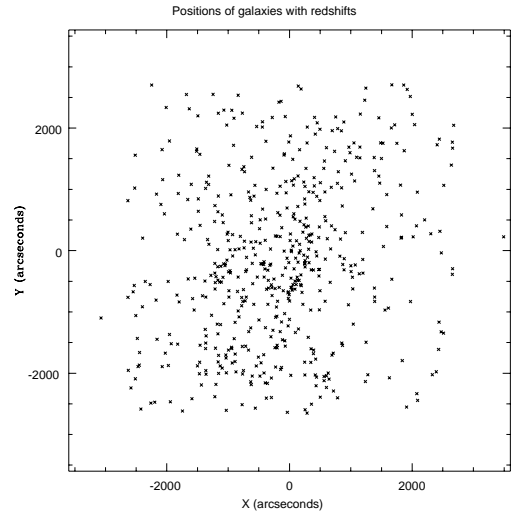,height=7cm}}
\caption[ ]{Spatial distribution of the 551 galaxies with redshifts}
\protect\label{xy}
\end{figure}

\begin{figure*}[htbp!]
\caption[ ]{Photograph of the MEFOS instrument.}
\protect\label{mefos}
\end{figure*}

Out of the 519 spectra obtained, we measured 421 reliable redshifts (the other
ones were discarded due to insufficient signal to noise). 
Our catalogue includes these spectra, plus those previously published
by Beers et al. (1991) and Malumuth et al. (1992). For galaxies observed twice,
we chose the redshift with the smallest error (usually the Beers et al. data).
The positions of the objects for which we obtained reliable spectra are
shown in Fig.~\ref{xy}. These positions are relative to the following cluster 
center: $\alpha_{2000.0} = 00^h41^{mn}51.90^s, \delta_{2000.0} = -09^\circ 18'17.0''$. This 
center was chosen to coincide with that of the diffuse X-ray gas component as 
defined by Pislar et al. (1997).

\subsection{Description of the MEFOS instrument}

MEFOS uses the big advantage of the prime focus for fibre
spectroscopy.  The 3.6m ESO telescope has a prime focus triplet
corrector delivering a field of one degree, the biggest at that time
for a 4m class telescope.  This will no longer be the case once the
2dF project at the AAT reaches completion in a very near future.  The
focal ratio is F/3.14, well suited for fibre light input, leading to
negligible focal ratio degradation. MEFOS (Gu\'erin et al. 1993) is
sitting on the red triplet corrector and is made of 30 arms that 
sweep the 20~cm diameter (one degree) field.

Fig.~\ref{mefos} shows the general arm
display. In fact, only 29 arms are positioned on astronomical objects, one
arm being used for guiding. The arms are displayed around the field as
``fishermen-around-the-pond''. The arms are moving radially and in rotation,
in such a way that each arm is acting in a 15 degree triangle with its
summit at the arm rotation axis and its base in the centre of the field.
So, all arms may access an object in the centre of the field and only one
can reach an object at the field periphery. This situation changes
gradually from the centre to the edge of the field. Each arm has its
individual electronic slave board and all the instrument is under control
of a PC computer, independent from the Telescope Control System (TCS). The
arm tips carry two fibres 1~arcmin apart, each one intercepting 2.5~arcsec on 
the sky. One is used for the object, and the
second one for the sky recording, and both go down to the spectrograph.
Object and sky can be exchanged; this allows to cancel the fibre transmission 
effects. 

\begin{figure}[htbp!]
\caption[ ]{Photograph of a galaxy field seen by the 29 windows on the CCD,
showing the position of the selected galaxies on the image bundles; this
allows to place the spectroscopic fibers accurately on each galaxy.}
\protect\label{mefosima}
\end{figure}

Coupled firmly to the arm tip is inserted an image conducting fibre 
bundle, that covers an area of 36$\times$36~arcsec$^2$ on the sky. All the 
image bundles are projected on a single Thomson 1024$\times$1024 thick CCD, 
Peltier cooled, connected to the same PC as the one driving the arms. 
Fig.~\ref{mefosima} shows a
galaxy field as seen by the 29 windows on the CCD, corresponding to the
arm image bundles set on the object coordinates. This procedure, in
opposition to blind positioning, is the only one, to our knowledge, that
shows the objects on which the spectral fibre will be placed in a second step. 
By analyzing the real position of the object in the image fibre, and
knowing the relative position of this image bundle and its connected
spectral fibre, a precise offset is computed and the arm is sent to its
working position. This offset takes care of all imprecisions due to the
telescope, the instrument and the coordinate inaccuracy. Given the poor
pointing of the telescope and the fact that the corrector and the
instrument are frequently dismounted, blind positioning would be
extremely dangerous. The positioning accuracy, as measured on  stellar
sources, is 0.2~arcsec.
In the present stage, the spectral fibres, 135~mm in diameter and 21~m
long, are going down from the prime focus to the Cassegrain, where the B\&C
ESO spectrograph is located. This spectrograph is fitted with a F/3
collimator to match the fibre output beam aperture, it has a set of
reflection gratings and a Tek 512 x 512 thin CCD.

\subsection{Data reduction}

The spectra were reduced using the IRAF software. The frames were bias and 
flat field corrected in the usual way. Velocities were measured by
cross-correlating the observed spectra with different templates~: a spectrum of
M31 (kindly provided by J. Perea) at a velocity of $-300$~km~s$^{-1}$, and
stellar spectra of the standard stars HD~24331 and HD~48381, which were each
observed every night during our 1994 run.
The cross-correlation technique is that described by Tonry \& Davis (1979) and 
implemented in the XCSAO task of the  RVSAO package in IRAF (Kurtz et al. 
1991).

The positions of emission-lines, when present, were measured by
fitting each line with a gaussian. 

All the spectra were reduced by the same person (F.D.) in a homogeneous way.
Redshifts of insufficient quality were discarded.

\section{Quality of the data}

\begin{figure}
\centerline{\psfig{figure=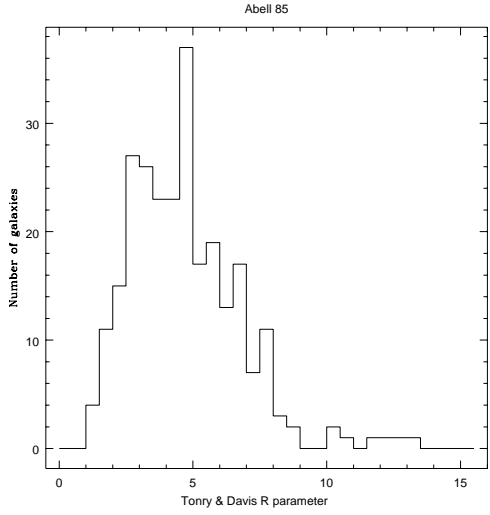,height=7cm}}
\caption[ ]{Distribution of the Tonry \& Davis R parameter given by the 
cross-correlation measure on absorption lines.}
\protect\label{Rparam}
\end{figure}

We classified our redshifts from 1 (best) to 3, according to their quality.
For galaxies with absorption lines:  
spectra of quality 1 have at least three lines clearly visible;
spectra of quality 2 have two; and 
spectra of quality 3 have only one.
The signal to noise parameter R given by the cross-correlation measure is
also given in Table~3. A histogram of this quantity is shown in
Fig.~\ref{Rparam}. 

For galaxies with emission lines: spectra of quality 1 have all
[OII]$\lambda$3727, H$\beta$, and [OIII]$\lambda\lambda$4959-5007
lines clearly visible; spectra of quality 2 have at least two emission
lines, and spectra of quality 3 have only one (usually
[OII]$\lambda$3727); in that case, the identification of the emission
line had to be confirmed by the shape of the continuum. The
identification of a single line was made possible by the fact that all
exposures were doubled, so we could remove cosmic rays and check that
the emission line was indeed present in both spectra.  Notice the high
number of redshifts (100) obtained from emission lines.

\begin{figure}
\centerline{\psfig{figure=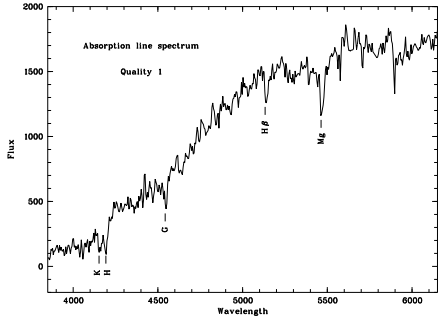,height=6.5cm,angle=-90}}
\caption[ ]{Typical absorption line spectrum of quality 1 (best).}
\protect\label{spec1}
\end{figure}

\begin{figure}
\centerline{\psfig{figure=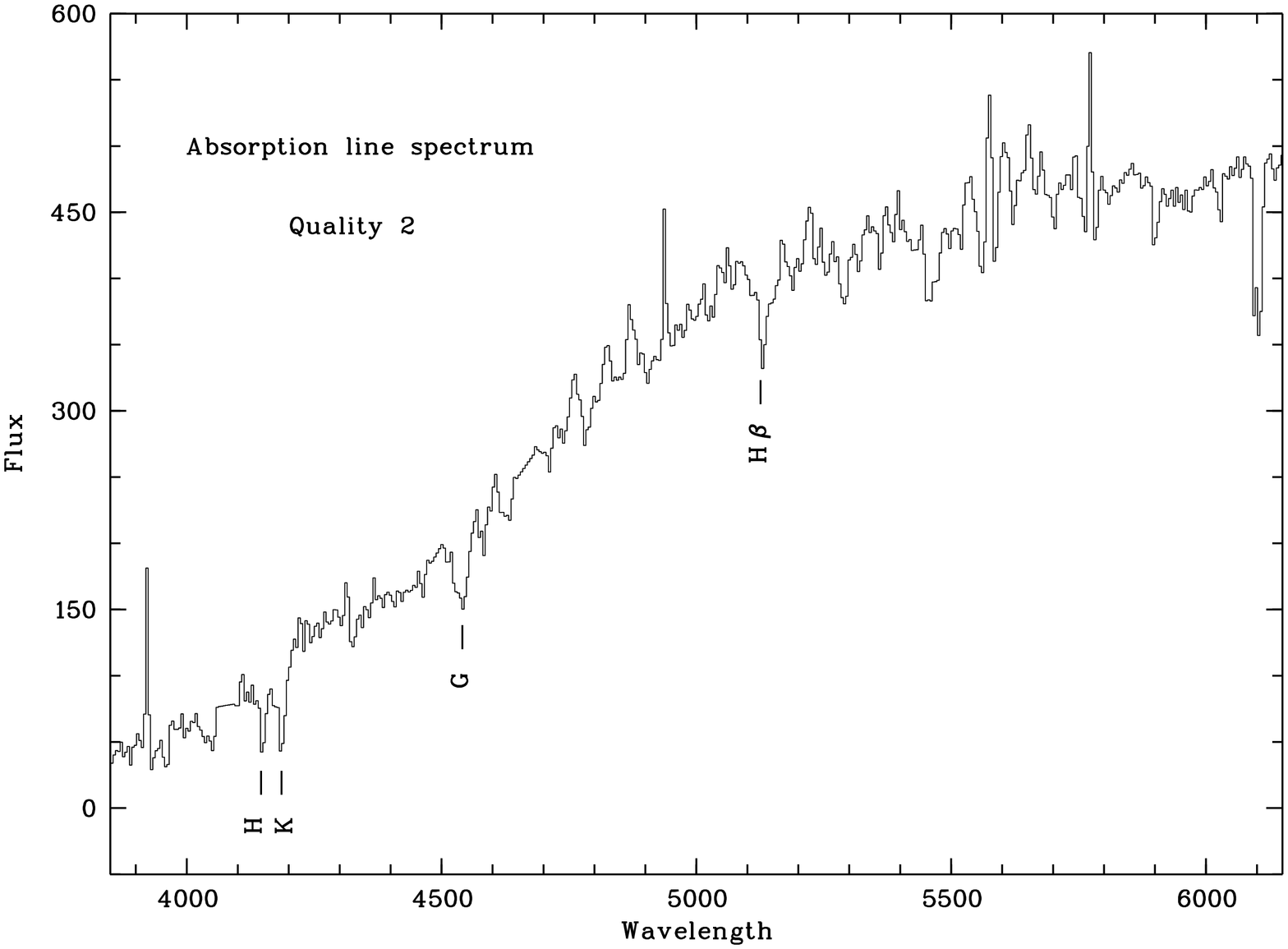,height=6.5cm,angle=-90}}
\caption[ ]{Typical absorption line spectrum of quality 2.}
\protect\label{spec2}
\end{figure}

\begin{figure}
\centerline{\psfig{figure=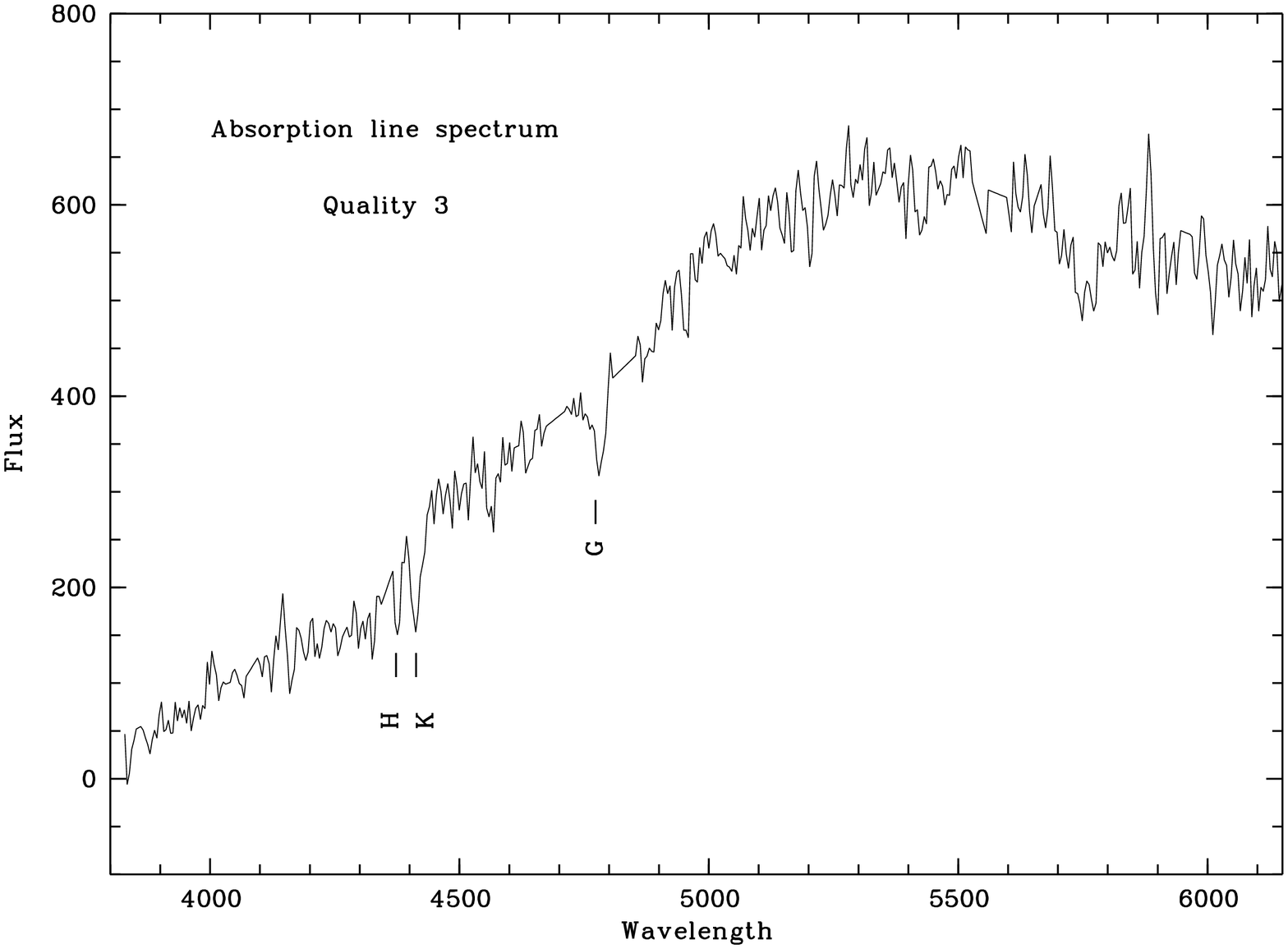,height=6.5cm,angle=-90}}
\caption[ ]{Typical absorption line spectrum of quality 3.}
\protect\label{spec3}
\end{figure}

\begin{figure}
\centerline{\psfig{figure=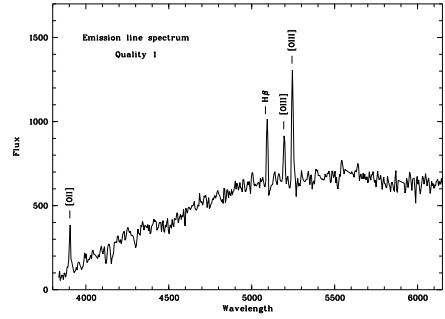,height=6.5cm,angle=-90}}
\caption[ ]{Typical emission line spectrum of quality 1 (best).}
\protect\label{spec4}
\end{figure}

\begin{figure}
\centerline{\psfig{figure=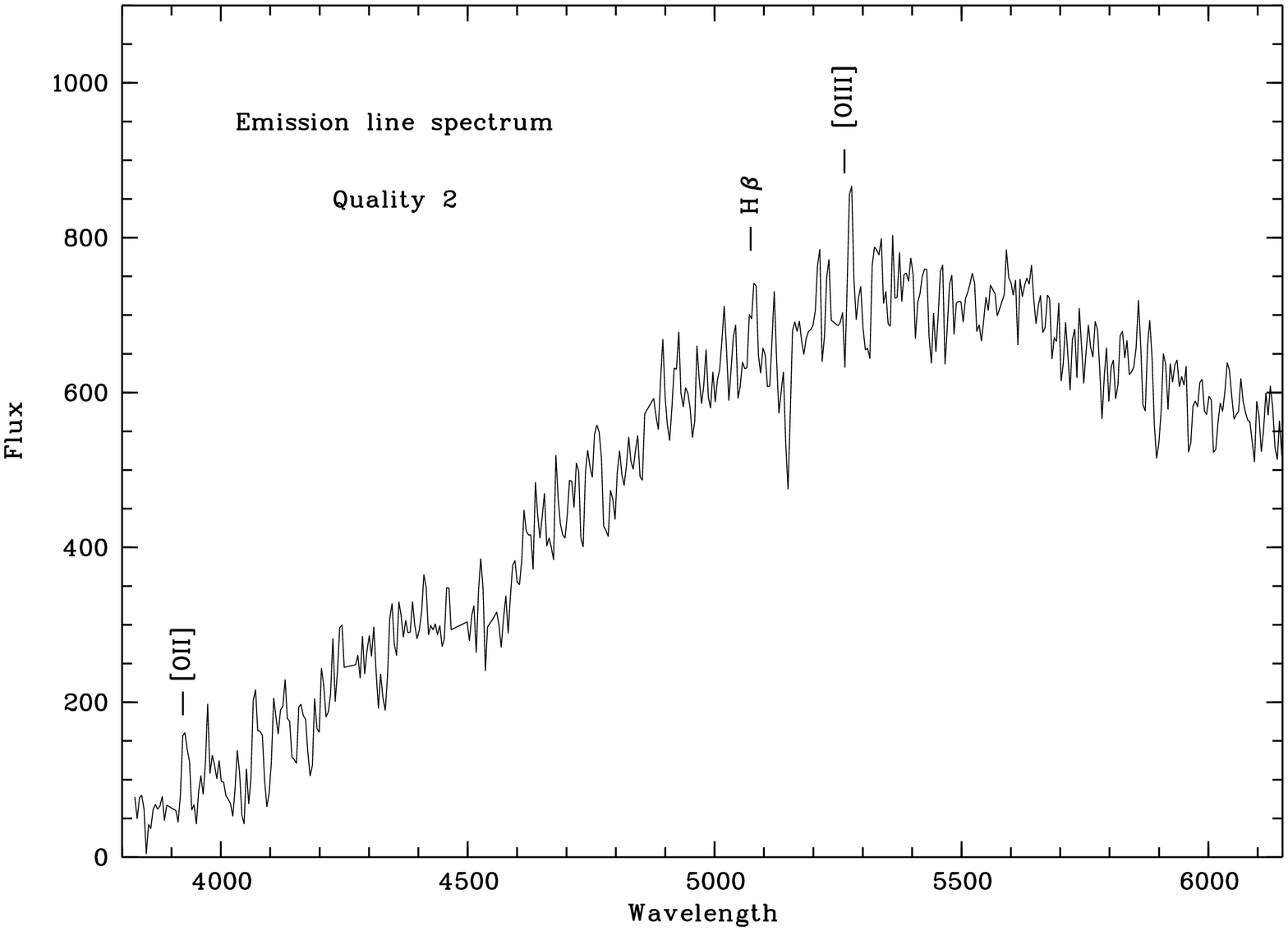,height=6.5cm,angle=-90}}
\caption[ ]{Typical emission line spectrum of quality 2.}
\protect\label{spec5}
\end{figure}

\begin{figure}
\centerline{\psfig{figure=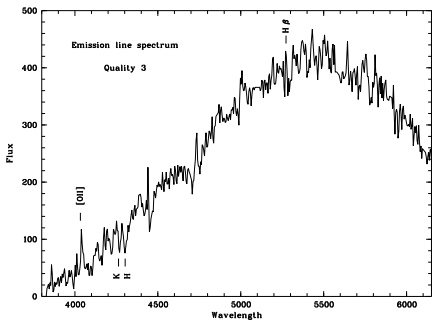,height=6.5cm,angle=-90}}
\caption[ ]{Typical emission line spectrum of quality 3.}
\protect\label{spec6}
\end{figure}

Typical spectra of various qualities are displayed in Figs.~\ref{spec1} to
\ref{spec6}.

\begin{figure}
\centerline{\psfig{figure=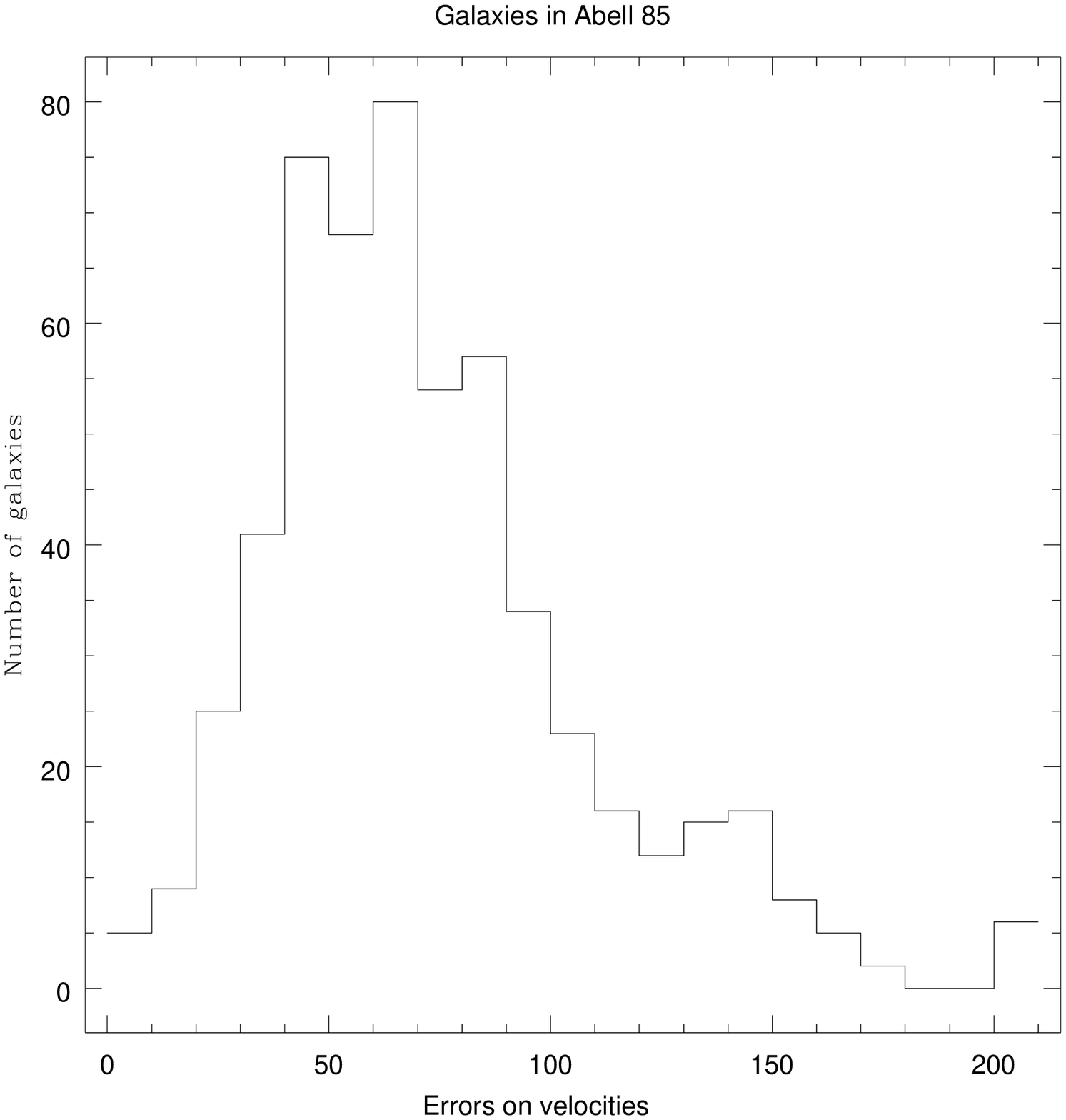,height=7cm}}
\caption[ ]{Distribution of internal errors on velocities derived from
absorption lines.}
\protect\label{errorvabs}
\end{figure}

\begin{figure}
\centerline{\psfig{figure=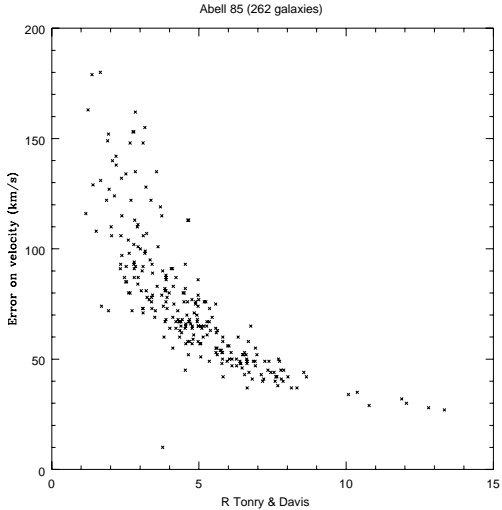,height=7cm}}
\caption[ ]{Relation between the Tonry \& Davis R parameter and the
errors on velocities derived from absorption lines.}
\protect\label{RTDerr}
\end{figure}

\begin{figure}
\centerline{\psfig{figure=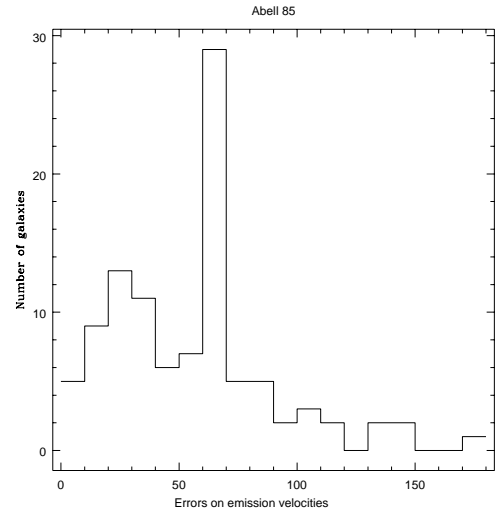,height=7cm}}
\caption[ ]{Distribution of internal errors on velocities derived from 
emission lines.}
\protect\label{errorvem}
\end{figure}

\begin{table*}
\caption{Comparison of galaxy velocities measured by us (DFLS) to those in the 
literature }
\begin{tabular}{rrrrrrr}
\hline
DFLS~~~~~~~~ & DFLS & DFLS & Beers~~~~~~~~  & Beers~~~~~~~~ & Beers & Difference \\
position~~~~~~~ & velocity & error & position~~~~~~~ & velocity & error & \\
\hline
0 41 28.112, -9 13 51.49 & 14102 & 92 & 0 41 27.202, -9 13 44.11 & 14233 & 46 & -131\\
0 41 28.112, -9 13 51.49 & 14102 & 92 & 0 41 29.000, -9 14 00.14 & 14022 & 33 &  80 \\
0 41 50.396, -9 18 09.48 & 16719 & 81 & 0 41 49.767, -9 18 33.40 & 16536 & 44 & 183 \\
0 41 50.396, -9 18 09.48 & 16719 & 81 & 0 41 50.070, -9 18 10.41 & 16734 & 48 & -15 \\
0 41 50.157, -9 25 47.54 & 17224 & 78 & 0 41 50.240, -9 25 47.41 & 17360 & 69 & -136 \\
0 41 53.400, -9 29 39.10 & 15331 & 48 & 0 41 53.625, -9 29 45.45 & 15392 & 34 & -61 \\
0 42 12.944, -9 17 50.97 & 19950 & 75 & 0 42 12.750, -9 17 50.70 & 19758 & 59 & 192 \\
\hline
DFLS~~~~~~~~ & DFLS & DFLS & Malumuth~~~~~~ & Malumuth & Malumuth & Difference \\
position~~~~~~~ & velocity & error & position~~~~~~~ & velocity & error & \\
\hline
0 40 31.650, -9 13 20.02 & 14102 & 42 & 0 40 31.550, -9 13 20.41 & 15410 & 91 & -1308 \\
0 40 43.550, -8 57 27.33 & 16530 & 48 & 0 40 43.550, -8 57 27.33 & 16341 & 32 &   189 \\
0 41 21.950, -9 03 31.18 & 16646 & 43 & 0 41 21.950, -9 03 31.18 & 16462 & 77 &   184 \\
0 42 12.940, -9 17 50.85 & 19950 & 67 & 0 42 12.940, -9 17 50.85 & 19930 & 45 &   -20 \\
0 42 35.640, -9 03 48.28 & 16287 & 43 & 0 42 35.640, -9 03 48.28 & 16301 & 74 &   -14 \\
0 43 10.730, -9 26 18.43 & 28724 & 30 & 0 43 10.730, -9 26 18.43 & 28890 & 93 &  -166 \\
\hline
Malumuth~~~~~~ & Malumuth & Malumuth & Beers~~~~~~~~ & Beers & Beers & Difference \\
position~~~~~~~ & velocity & error & position~~~~~~~ & velocity & error & \\
\hline
0 38 30.1, -9 29 01 & 16256 & 96 & 0 38 30.0, -9 29 02 & 16213 & 48 & 43 \\
0 38 33.7, -9 27 60 & 16617 & 83 & 0 38 33.7, -9 27 59 & 16725 & 64 &-108 \\
0 38 55.2, -9 30 09 & 14178 &109 & 0 38 55.4, -9 30 11 & 14233 & 46 & -55 \\
0 38 56.9, -9 30 25 & 16530 & 61 & 0 38 57.2, -9 30 27 & 14022 & 33 &2508 \\
0 38 58.4, -9 32 13 & 13393 & 26 & 0 38 58.6, -9 32 15 & 13429 & 28 & -36 \\
0 38 58.6, -9 30 33 & 16241 & 96 & 0 38 58.7, -9 30 35 & 16328 & 46 & -87 \\
0 39 00.1, -9 36 29 & 13811 & 56 & 0 39 00.2, -9 36 30 & 13781 & 41 &  30  \\
0 39 01.7, -9 25 51 & 18141 & 59 & 0 39 01.6, -9 25 52 & 18154 & 39 & -13 \\
0 39 02.9, -9 38 17 & 14125 & 46 & 0 39 03.3, -9 38 19 & 14234 & 42 &-109 \\
0 39 11.2, -9 42 50 & 16792 &107 & 0 39 11.3, -9 42 49 & 16886 & 35 & -94 \\
0 39 16.4, -9 33 30 & 15851 & 22 & 0 39 16.6, -9 33 31 & 15912 & 69 & -61 \\
0 39 18.6, -9 34 38 & 16447 &126 & 0 39 18.3, -9 34 37 & 16734 & 48 &-287 \\
0 39 18.4, -9 42 15 & 17174 &118 & 0 39 18.5, -9 42 14 & 17360 & 69 &-186 \\
0 39 20.4, -9 46 42 & 17054 & 51 & 0 39 20.5, -9 46 43 & 17164 & 33 &-110  \\
0 39 21.7, -9 46 13 & 15181 & 46 & 0 39 21.9, -9 46 12 & 15392 & 34 &-241 \\
0 39 41.1, -9 34 17 & 19930 & 45 & 0 39 41.0, -9 34 17 & 19758 & 59 & 172 \\ 
0 40 01.9, -9 27 05 & 16891 & 99 & 0 40 02.0, -9 27 07 & 16761 & 58 & 130 \\
0 40 22.9, -9 30 20 & 22910 & 51 & 0 40 22.8, -9 30 16 &  5230 & 59 &17680 \\
\hline

\end{tabular}
\end{table*}

In order to check the intrinsic quality of our velocity measurements, two
velocity standard stars from the Maurice et al. (1984) list were observed each 
night. The errors, derived by cross-correlating the star spectra to the
spectrum of M31, range (from night to night) from $\pm$16
to $\pm 23$~km~s$^{-1}$ for HD~24331, and from $\pm$17 to 
$\pm 43$~km~s$^{-1}$ HD~48381. The mean internal
error on velocities derived from the accuracy in the wavelength
calibration is 66~km~s$^{-1}$. The distribution of
errors on velocities is displayed in Figs.~\ref{errorvabs}
and \ref{errorvem} for absorption and emission line measurements
respectively (these histograms only include the galaxies that we
observed).  For absorption lines, the correlation between the Tonry \& Davis
R parameter and the error on the velocity is displayed in Fig.~\ref{RTDerr}
for the 262 galaxies for which we had both quantities. 
For emission line measurements, the errors on velocities
were estimated from the dispersion on the velocities derived from the
various emission lines present. When only one emission line was
present we averaged the emission and absorption line redshifts
whenever possible; if no reliable absorption line redshift was
available, we estimated the internal error on a single emission line
to be the intrinsic value of 66~km~s$^{-1}$ mentioned above.

In order to test the agreement of our redshifts with those of previous
surveys, we reobserved 7 and 6 galaxies from the velocity catalogues
by Beers et al. (1991) and Malumuth et al. (1992) respectively. The
comparison of these various measurements are given in Table~1.  Except
for a few totally discrepant values, which are probably due to galaxy
misidentifications, our values agree with those of the literature,
with BWT mean differences of +8 and +35~\kms between our values and
the Beers and Malumuth samples respectively, and corresponding BWT
dispersions of 132 and 156~km~s$^{-1}$ (for the comparison with the
Beers data, we kept the two first of our objects in the above Table
which gave the smallest velocity difference with the Beers
data). These values are comparable to the BWT mean ($-71$~\kms ) and
BWT dispersion (125~\kms ) that we estimate between the Beers and
Malumuth samples.  However, the number of objects in common is of
course small and may pervert statistics.

To check the consistency of our calibrations between the two observing runs, we
reobserved in 1995 four galaxies already observed in 1994; the agreement between 
the various values is better than 100~\kms, confirming that the mean
error on our velocities is smaller than 100~\kms. 

The final redshifts given in the catalogue are those derived from 
the cross-correlation with M31, since this template gave the best results.
A correction was applied to obtain heliocentric velocities.

\begin{figure}
\centerline{\psfig{figure=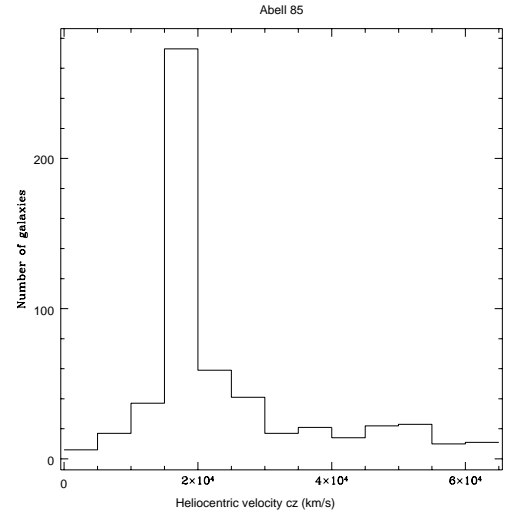,height=7cm}}
\caption[ ]{Velocity histogram of all observed galaxies}
\protect\label{histov}
\end{figure}

The histogram of all the velocities in the catalogue is displayed in 
Fig.~\ref{histov}. It will be discussed in detail in a companion paper
(Durret et al. in preparation).

\begin{table*}
\caption{Completeness of the redshift catalogue for different field 
diameters within a limiting magnitude R$\leq$18. Numbers in parentheses
indicate the absolute numbers of galaxies with and without redshifts
respectively. Note that 1000'' correspond to 157~kpc with
H$_\circ$=50~km~s$^{-1}$~Mpc$^{-1}$.}

\begin{tabular}{lccccc}
\hline
          &  &      & & & \\
Limiting  & 700  & 1000 & 1500 & 2000 & 2500 \\
diameter  ('')   &      & & & \\
          &  &      & & & \\
\hline
          &  &      & & & \\
Completeness & 88.5\% (85/96) & 91.9\% (137/149) & 87.6\% (219/250) & 
84.7\% (316/373) & 77.4\% (384/496) \\
          &  &      & & & \\
\hline
\end{tabular}
\protect\label{compl}
\end{table*} 

We have estimated the completeness of the spectroscopic catalogue presented 
here by comparing the numbers of galaxies with redshifts to the total
number of galaxies from our photographic plate catalogue. Results are
shown in Table~\ref{compl}. 

\section{The catalogues}

\begin{figure}
\centerline{\psfig{figure=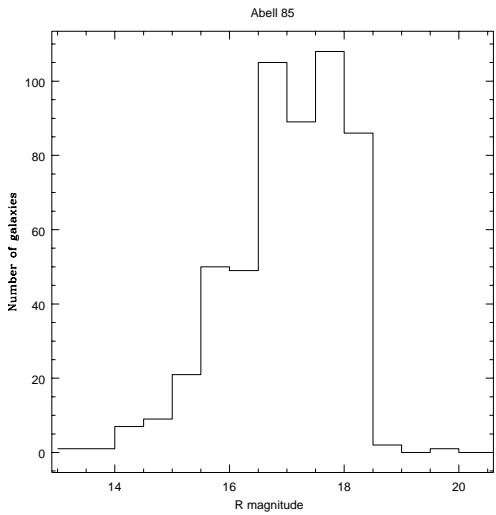,height=7cm}}
\caption[ ]{Magnitude histogram in the R band of all galaxies with measured
velocities}
\protect\label{maggal}
\end{figure}

\begin{figure}
\centerline{\psfig{figure=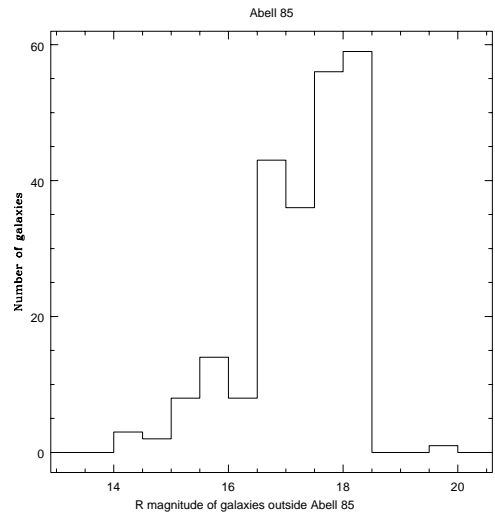,height=7cm}}
\caption[ ]{Magnitude histogram in the R band of galaxies with measured
velocities outside the velocity range defined for the cluster.}
\protect\label{maggalout}
\end{figure}

\begin{figure}
\centerline{\psfig{figure=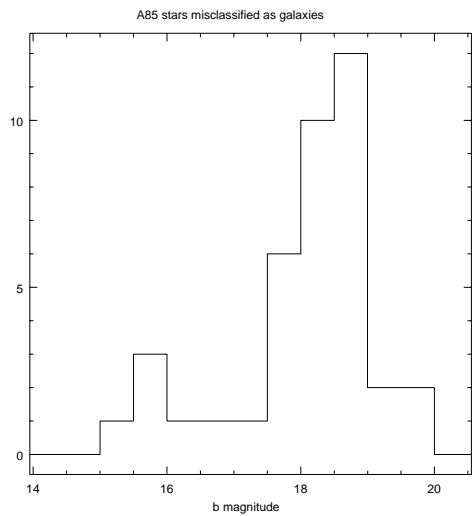,height=7cm}}
\caption[ ]{Magnitude histogram in the \bj\ band of the 39 stars misclassified 
as galaxies}
\protect\label{magstars}
\end{figure}

Note that out of the 421 reliable redshifts that we obtained, 39 were
those of stars misclassified as galaxies in our photometric plate
catalogue. In order to show that most misclassifications were made on
faint objects, we show in Figs.~\ref{maggal} and \ref{magstars} the
magnitude histograms for the galaxies with measured velocities and
those for the 39 stars. Although this number may seem large, it
corresponds to a percentage of contamination only 5\%\ for \bj $\leq
18.5$, in agreement with the verification made by eye on a portion of
the catalogue.  The contamination by foreground and background
galaxies is illustrated in Fig.~\ref{maggalout}, which shows the histogram
of the R magnitudes for the galaxies outside the velocity range
13350-2000~km/s corresponding to \a85.

\begin{table}
\caption{Coordinates of the 39 stars misclassified as galaxies.}
\begin{tabular}{rrr}
\hline
Running & $\alpha$~~~ & $\delta$~~~ \\
number &  \multicolumn{2}{c}{(J2000.0)} \\
\hline
 1 & 0 38 53.96 &  -8 54 47.39 \\
 2 & 0 38 54.64 &  -8 54 58.72 \\
 3 & 0 39 04.83 &  -9 34 29.83 \\
 4 & 0 39 19.77 &  -8 33 45.12 \\
 5 & 0 39 21.33 &  -9 43 02.95 \\
 6 & 0 40 19.95 &  -9 12 56.30 \\
 7 & 0 40 20.74 &  -9 37 48.60 \\
 8 & 0 40 24.95 &  -8 59 37.11 \\
 9 & 0 40 35.43 &  -9 53 18.82 \\
10 & 0 40 47.65 &  -8 54 29.04 \\
11 & 0 40 47.76 &  -8 40 06.64 \\
12 & 0 40 50.14 &  -9 29 24.14 \\
13 & 0 41 10.53 &  -9 49 50.17 \\
14 & 0 41 42.94 &  -9 19 14.51 \\
15 & 0 41 54.45 &  -8 44 30.03 \\
16 & 0 41 56.15 &  -8 53 15.83 \\
17 & 0 42 02.95 &  -8 38 03.54 \\
18 & 0 42 07.23 &  -9 52 22.05 \\
19 & 0 42 16.94 &  -9 11 10.46 \\
20 & 0 42 17.13 &  -9 33 25.16 \\
21 & 0 42 18.31 &  -9 59 02.14 \\
22 & 0 42 21.05 &  -8 58 20.66 \\
23 & 0 42 25.73 &  -9 44 56.37 \\
24 & 0 42 26.24 &  -9 15 19.57 \\
25 & 0 42 56.33 &  -9 29 50.41 \\
26 & 0 43 05.75 &  -8 43 33.93 \\
27 & 0 43 10.04 &  -9 02 29.83 \\
28 & 0 43 16.14 &  -8 49 49.14 \\
29 & 0 43 18.14 &  -9 13 07.34 \\
30 & 0 43 19.12 &  -9 48 52.45 \\
31 & 0 43 20.41 & -10 01 11.80 \\
32 & 0 43 23.33 &  -9 18 33.75 \\
33 & 0 44 05.91 &  -9 56 30.21 \\
34 & 0 44 06.75 &  -8 32 35.11 \\
35 & 0 44 27.94 &  -8 59 08.44 \\
36 & 0 44 37.74 &  -8 41 39.96 \\
37 & 0 44 38.41 &  -9 54 48.36 \\
38 & 0 44 43.31 &  -9 57 11.87 \\
39 & 0 44 44.24 &  -8 48 49.47 \\
\hline
\end{tabular}
\protect\label{stars}
\end{table} 

The coordinates (equinox 2000.0) of the 39 stars misclassified as
galaxies are given in Table~\ref{stars}, to avoid further observations
of these objects in galaxy surveys.

The velocity data for the galaxies in the field of ABCG~85 are given 
in Table~4 (available in electronic form only). The signification of the 
columns is the following:\\
(1) Running number; \\
(2) to (4): right ascension (equinox 2000.0); \\
(5) to (7): declination (equinox 2000.0); \\
(8): heliocentric velocity (cz) in \kms; \\
(9): error on the velocity in \kms; for velocities derived from absorption 
lines, the error is either that stated in the literature or, for our own 
measurements, that given by the RVSAO IRAF package;  for velocities derived 
from several emission lines, the error was estimated from the dispersion
between the velocities derived from the different emission lines; when 
only one emission line was present and an absorption line redshift was
obtainable, the error on the velocity was taken to be the dispersion
between both measurements; finally, when only one emission line was present 
and no absorption line redshift was available, the error on the velocity
was taken to be the mean internal velocity error; \\
(10): Tonry \& Davis R parameter;
(11): label indicating the means of determination of the redshift: 0=derived 
from absorption lines, 1=derived from emission lines; \\
(12): label indicating the quality of the data, from 1 (best) to 3 (0 for
data taken from the literature); \\
(13): label indicating the origin of the data: 1=Malumuth et al. (1992),
2=Beers et al. (1991), 3=our data; \\
(14) and (15): X and Y positions in arcseconds relative to the center
assumed to have coordinates \\
$\alpha = 00^h41^{mn}51.90^s, \delta = -09^\circ 18'17.0''$ (equinox 2000.0); 
\\
(16) distance to the cluster center in arcseconds;\\
(17) to (19): magnitudes in the b$_{\rm J}$, V and R bands respectively; \\
(20): reference to the galaxy number in the Beers et al. 1991 (B) and
Malumuth et al. 1992 (M) catalogues, as well as reference to the name given 
to six galaxies identified as X-ray sources by Pislar et al. 1997
(their Table~1).\\

\section{Conclusions} 

Our photometric catalogues, including both the large field catalogue
obtained by scanning a photographic plate and the small CCD field
catalogue, will be published in a companion paper (Slezak et
al. 1997).  All these optical data are used to give an interpretation
of the properties of \a85 (Durret et al. in preparation).  They will
be compared to the results already obtained from X-ray data in
Papers~I and II (Pislar et al. 1997, Lima--Neto et al.  1997).

\acknowledgements {We are very grateful to Andr\'ee Fernandez for her help
during the preparation of the first observing run, to Paul Stein for helping 
us obtain a catalogue of guiding stars and to Cl\'audia Mendes de Oliveira for 
her cheerful and highly competent assistance at the telescope. We thank the
anonymous referee for many useful comments. CL is fully supported by the 
BD/2772/93RM grant attributed by JNICT, Portugal.}

\end{document}